\documentclass[aps,prb,twocolumn,superscriptaddress,showpacs]{revtex4}
\usepackage[T1]{fontenc}
\usepackage[latin9]{inputenc}
\usepackage{amsmath}
\usepackage{amssymb}
\usepackage{graphicx}
\usepackage{esint}
\usepackage{bm}
\usepackage{multirow}

\renewcommand{\vec}[1]{\bm{#1}}
\newcommand{\dif}{\mathrm{d}}
\DeclareMathAlphabet{\mathsfsl}{OT1}{cmss}{m}{sl}

\begin{document}

\title{Floquet Majorana fermions in driven hexagonal lattice systems}

\author{Zi-bo Wang}
\affiliation{International Center for Quantum Materials, School of Physics, Peking
University, Beijing 100871, China}
\affiliation{Collaborative Innovation Center of Quantum Matter, Beijing 100871, China}

\author{Hua Jiang}
\affiliation{Department of Physics, Soochow University, Suzhou 215006, China}

\author{Haiwen Liu}
\thanks{\texttt{haiwen.liu@pku.edu.cn}}
\affiliation{International Center for Quantum Materials, School of Physics, Peking
University, Beijing 100871, China}
\affiliation{Collaborative Innovation Center of Quantum Matter, Beijing 100871, China}

\author{X. C. Xie}
\affiliation{International Center for Quantum Materials, School of Physics, Peking
University, Beijing 100871, China}
\affiliation{Collaborative Innovation Center of Quantum Matter, Beijing 100871, China}

\date{\today}

\begin{abstract}
We propose Floquet chiral topological superconducting systems hosting Floquet Majorana fermions, which consist of hexagonal lattices in proximity to superconductors with shining circularly polarized light. Specially for bilayer graphene system, we demonstrate that there exist three topological phases determined by the system parameters, namely, the amplitude and frequency of the induced light. The number of chiral Floquet Majorana edge states is confirmed by calculating the Chern number analytically and the energy spectrum in ribbon geometry. Moreover, this proposal is generalized to other hexagonal lattice systems, such as the monolayer graphene and silicene. Notably, the parameter range of induced light to achieve the chiral Floquet Majorana edge states is experimentally feasible, and the corresponding Floquet Majorana fermions can be probed based on differential conductance using the scanning tunneling spectroscopy.
\end{abstract}

\date{\today}

\maketitle

\section{Introduction}
Majorana fermions (MFs) are ``real-valued'' fermion modes, which are their own anti-particles in contrast to other ``complex'' fermions. A pair of widely spatial seperated MF bound states corresponds to zero-energy modes of Bogoliubov\(\text{-}\)de Gennes (BDG) Hamiltonian, and forms non-Abelian quasiparticle states immune to local decoherence, which suit for potential applications in quantum computation.\cite{Frank Wilczek,Chetan Nayak} MFs are proposed to emerge in kinds of 2D systems, such as the surface states of topological insulators proximated to superconductors,\cite{Liang} the \(\nu=5/2\) fractional quantum Hall(FQH) states,\cite{Gregory Moore} quantum anomalous hall(QAH) states in proximity to superconductors\cite{Xiao-Liang}, non-centrosymmetric superconductors\cite{Masatoshi}, \textit{etc}. Moreover, various 1D systems are also proposed to host MFs.\cite{Roman M. Lutchyn,Oreg2010} Importantly, high magnetic field, which breaks the time reversal symmetry, is essential for the appearance of MFs in most of these strategies. However, the high magnetic field might easily break the superconducting layer, which imposes restrictions on the realization of MFs in experiments.

Recently, topological insulators driven by external time-dependent perturbations have stimulated great scientific interests.\cite{Cayssol} Specifically, for the time-periodic driven systems, it is convenient to use Floquet theory to characterize their topological phases.\cite{F. H. M. Faisal,Han,Gomez-Leon} For example, with shinning spatial modulating irradiation on a semiconductor quantum well, which is initially in the trivial phase, topological nontrivial edge states can be induced in Floquet bands.\cite{Netanel, Yaniv Tenenbaum Katan} Additionally, with circularly polarized light, a nontrivial gap is opened and edge states emerge in hexagonal lattices, such as graphene and silicene.\cite{Zhenghao, Takuya Kitagawa1,Motohiko Ezawa} Importantly, Mikael et al., detected photonic Floquet edge states in photonic lattices experimentally.\cite{Mikael} And, with the help of time-resolved ARPES, Wang et al., have observed the Floquet states on the surface of a topological insulator.\cite{Y. H. Wang}

More recently, a new concept of Floquet Majorana fermions (FMFs) is brought in cold-atom systems, such as driven cold-atom quantum wires\cite{Liang Jiang,Andres A. Reynoso} and so on. Furthermore, FMFs can also encode quantum information when the periodic driving potential does not break fermion parity conservation.\cite{Dong}
However, there are few studies focusing on FMFs in condensed matter systems. Especially, with circularly polarized light, hexagonal lattices have been proved to be topological nontrival systems hosting Floquet edge states.\cite{Zhenghao, Takuya Kitagawa1,Motohiko Ezawa} One natural question is whether FMFs can exist in those systems. In fact, the hexagonal lattices possess unique advantages due to the existence of antiferromagnetic order.  In particular, the antiferromagnetic phase can coexist with superconductivity, breaking the time-reversal symmetry without external
magnetic field. Recently, it has been demonstrated that there exists the layer-antiferromagnetic(LAF) phase in bilayer graphene.\cite{B. E. Feldman, Y. Lemonik, Maxim Kharitonov,Yong Wang} And the antiferromagnetic phase has been also theoretically predicted existing in monolayer hexagonal lattices, such as graphene.\cite{A. H. Castro Neto} Considering these advantages, the hexagonal lattices provide a promising platform for realizing FMFs.

In this work, we study bilayer graphene in proximity to an s-wave superconductor with shinning circularly polarized light at high frequencies. We find that there exist three topological phases by tuning the physical parameters of the induced light: the amplitude \(A\) and the frequency \(\omega\). Based on the effective Hamiltonian, we calculate the Chern numbers--the number of chiral Floquet Majorana edge states to be 8, 4 and 0, respectively, corresponding to the three different topological phases. Moreover, the number of those edge states is also confirmed by solving energy spectra in the ribbon geometry. Notably, in narrow ribbons, due to the finite size confinement, there exist another two cases with edge states numbers 6 and 2. Furthermore, this proposal of generating chiral FMFs is generalized to other hexagonal lattices, such as the single-layer graphene and silicene.
In silicene, the \(\sqrt{3}\times\sqrt{3}R\) reconstruction hosts odd numbers of FMFs, which is important in realization
of the non-trival braiding statistics.
Finally, we propose a strategy to probe those Floquet Majorana edge states using the scanning tunneling spectroscopy (STS).

The rest of this paper is organized as follows. In Sec.II, we present the proposed system and derive the effective Hamiltonian (see details in App.A) describing the system. Then, in Sec.III, a topological phase diagram is obtained based on the Chern number calculated in App.C. Next, in Sec.IV, the chiral Floquet Majorana edge states are further confirmed with the numerical results
of the energy band structures in the ribbon geometry. In Sec.V, we extend aforementioned method for generating FMFs to
other kinds of the hexagonal lattices and propose a scheme to detect the FMFs.
Finally, a conclusion is presented in Sec.VI.

\section{Proposed device and model Hamiltonian}

\begin{figure}[h]
\includegraphics [width=\columnwidth, viewport=0 0 785 639, clip]{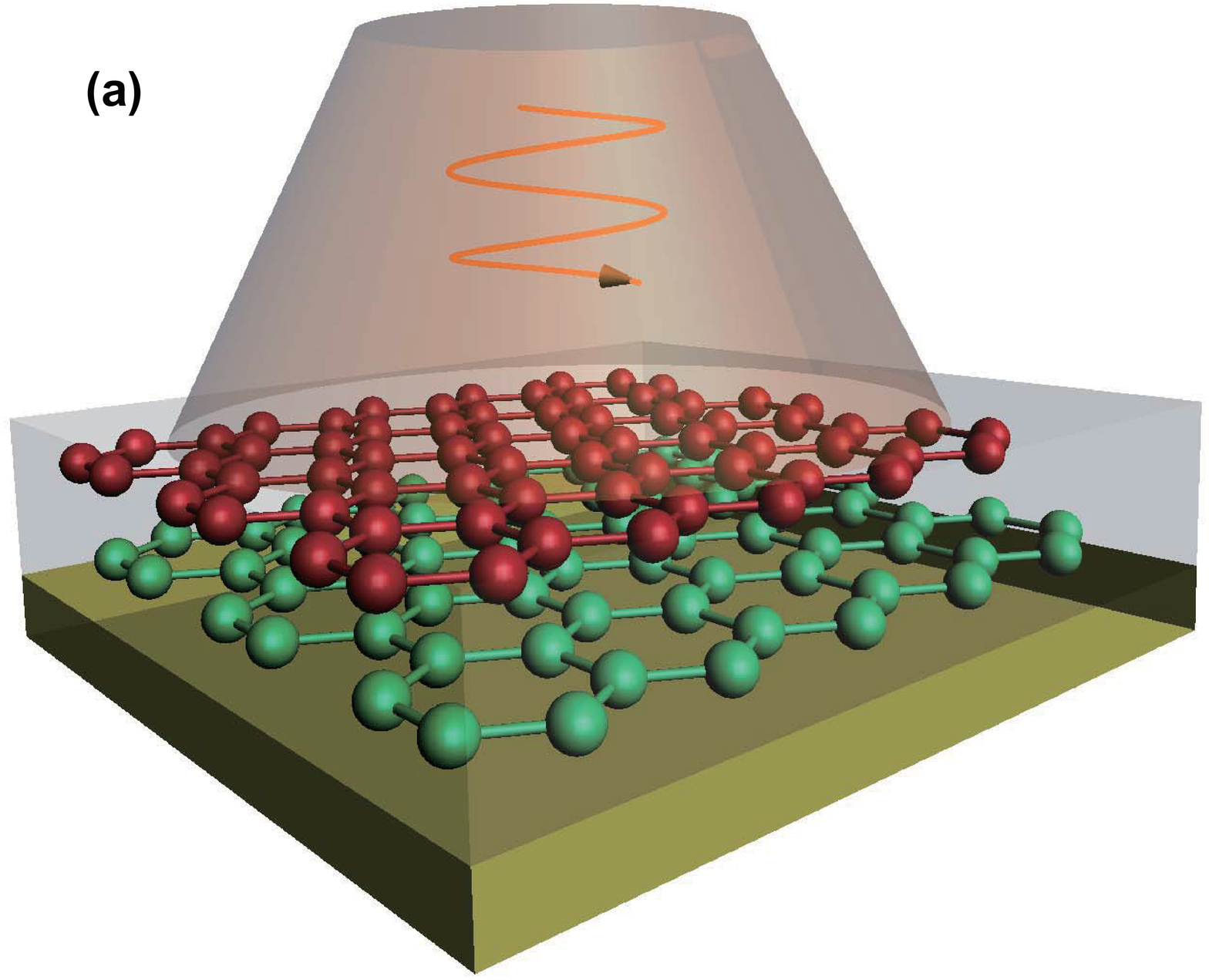}
\includegraphics [width=\columnwidth, viewport=0 0 827 350, clip]{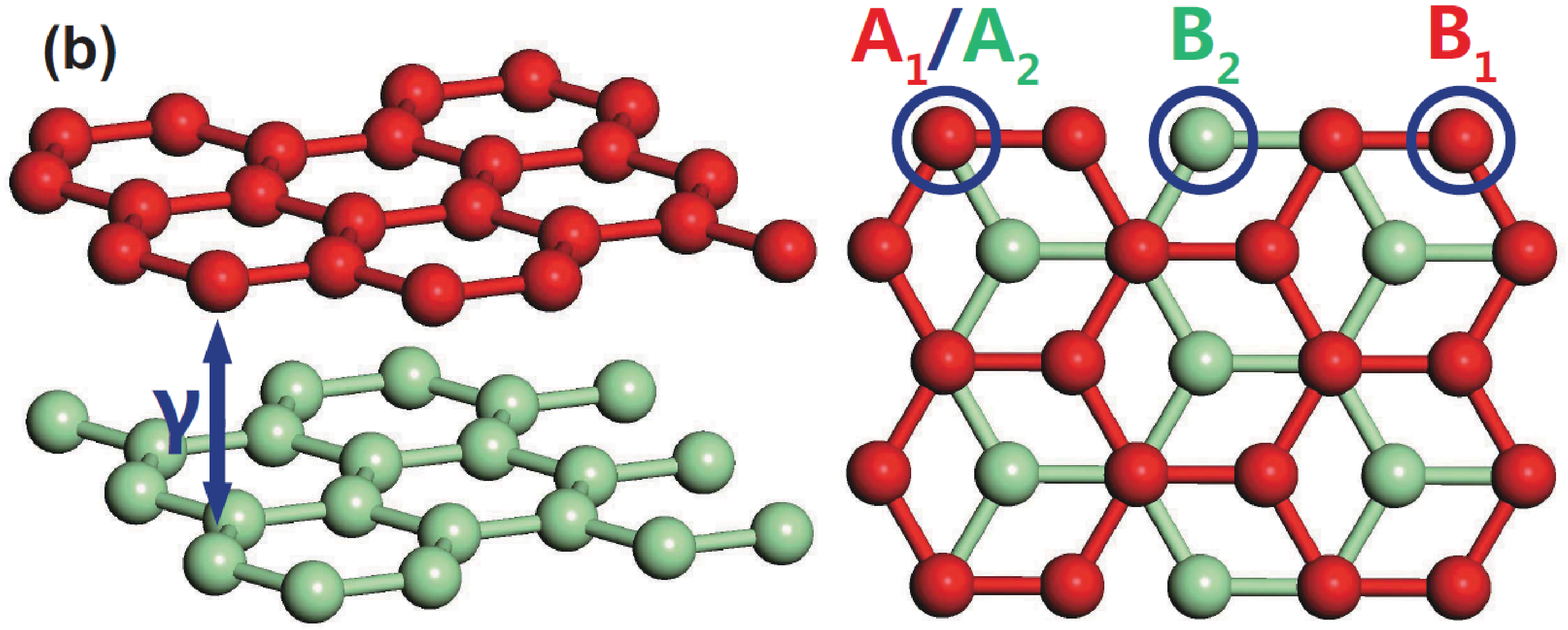}
\caption{(Color online) (a) is the schematic diagram of proposed device,
which consists of an s-wave superconductor layer at the bottom and a bilayer graphene
layer on top. A beam of circularly polarized light is projected  perpendicularly to the bilayer graphene plane with layer-antiferromagnetic effect, and the light induced nearest-neighbour hopping \(m\) is determined by the amplitude \(A\) and frequency \(\omega\) of induced light. (b) is the lattice structure of bilayer graphene.
Left panel: Side view of bilayer graphene, where
\(\gamma\) is the hopping energy between atom \(A_1\) and atom \(A_2\);
Right panel: Top-down view of the lattice structure,
in which \(A_{1(2)}\) and \(B_{1(2)}\) represent different sublattice.
In this figure, red and green balls stand for atoms in upper and lower layers, respectively.
}
\end{figure}

We first study the bilayer graphene system, due to the experimental confirmation of the layer-antiferromagnetic(LAF) phase in such system.\cite{B. E. Feldman, Y. Lemonik, Maxim Kharitonov,Yong Wang} Our proposed device is illustrated in Fig.1(a). The bilayer graphene with LAF is put in proximity to an s-wave superconductor.
Meanwhile, a beam of right-handed circularly polarized light is projected perpendicularly
to the plane of bilayer graphene (left-handed rotation can be discussed similarly).
Thus, the effective tight-binding Hamiltonian \(H_b\) can be written as:
\begin{eqnarray}
    H_b & = & -t\sum_{<ij>,s,\alpha}(a^\dagger_{i,s,\alpha}b_{j,s,\alpha}+H.c.) \nonumber\\
      &   & -\gamma\sum_{i,s}(a^\dagger_{i,s,1}a_{i,s,2}+H.c.) \nonumber\\
      &   & +\frac{im}{3\sqrt{3}}\sum_{\ll ij\gg,s,\alpha}
       [v_{ij}(a^\dagger_{i,s,\alpha}a_{j,s,\alpha}+b^\dagger_{i,s,\alpha}b_{j,s,\alpha})] \nonumber\\
      &   & +V\sum_{i,s}\ [s(b^\dagger_{i,s,1}b_{i,s,1}-b^\dagger_{i,s,2}b_{i,s,2})]
\end{eqnarray}
where \(a_{i,s,\alpha}(a^\dagger_{i,s,\alpha})\) annihilates (creates) an electron with spin \(s(s=\pm1)\) on site \(\vec{R}_i\) in layer \(\alpha(\alpha=1,2)\) on sublattice \(A\) (an equivalent definition is used for sublattice B), as shown in Fig.1(b). In \(H_b\), the first two terms denote intrinsic hopping in bilayer graphene, where \(t\) in the first term is the nearest-neighbor hopping energy, and \(\gamma\) in the second term stands for the hopping energy between atom \(A_1\) and \(A_2\) in different layers. We consider the off-resonant circularly polarized light case, in which no inter-band electronic transitions occur due to the prohibition of direct photon absorption. The off-resonant light modifies the electron via virtual photon absorption process, and results in a chiral next nearest-neighbour hopping (see App.A for details), where \(m=v_g^2\mathcal{A}^2/\omega\) denotes the next nearest-neighbor
hopping energy, and \(v_{ij}=+1 (-1)\) corresponds to clockwise (anti-clockwise) hopping. For graphene systems, the off-resonant condition requires photon energies larger than band width, which lies in the range of soft X-ray regime.\cite{Takuya Kitagawa1,Motohiko Ezawa} Recently, the circularly light induced gap has been detected in the surface state of topological insulator, which is consistent with the effective model.\cite{Y. H. Wang} The last term in Eq.1 stands for LAF, which is spontaneously generated by electron-electron interaction in bilayer graphene around half filling, and \(V\) is the LAF strength.\cite{B. E. Feldman, Y. Lemonik, Maxim Kharitonov,Yong Wang} The LAF and superconductivity come from different orgin, and may coexist in the interface graphene layers. Previously, the coexistent of antiferromagnetism and superconductivity has been investigated theoretically and experimentally.\cite{Kazuhiro Kuboki,A. I. Buzdin}   In general, the circularly polarized light gives rise to the nontrivial topology, and the LAF term lifts the spin degeneracy. The combination of these two effects leads to abundant topological phases in hexagonal lattices as discussed in the following parts.

Considering the proximity effect of an s-wave superconductor as shown in Fig.1(a), the Bogoliubov\(\text{-}\)de Gennes (BdG) Hamiltonian of the proposed system can be obtained in the momentum space:
\begin{eqnarray}
    H_{BDG}=
    \sum_{\vec{k}}\Psi_{\vec{k}}^\dagger
    \left(
     \begin{array}{cc}
       H_b(\vec{k}) & i\Delta I_4\otimes\sigma_y\\
       -i\Delta I_4\otimes\sigma_y   & -H_b^*(-\vec{k})
     \end{array}
    \right)\Psi_{\vec{k}}
\end{eqnarray}
where \(\Psi_{\vec{k}}=(a_{1\vec{k}\uparrow}, a_{1\vec{k}\downarrow}, b_{1\vec{k}\uparrow},
b_{1\vec{k}\downarrow}, a_{2\vec{k}\uparrow}, a_{2\vec{k}\downarrow}, b_{2\vec{k}\uparrow},
b_{2\vec{k}\downarrow},\\ a^\dagger_{1\vec{-k}\uparrow}, a^\dagger_{1\vec{-k}\downarrow}, b^\dagger_{1\vec{-k}\uparrow},
b^\dagger_{1\vec{-k}\downarrow}, a^\dagger_{2\vec{-k}\uparrow}, a^\dagger_{2\vec{-k}\downarrow}, b^\dagger_{2\vec{-k}\uparrow},
b^\dagger_{2\vec{-k}\downarrow})^T\), \(\Delta\) denotes the superconducting order parameter and \(I_4=diag\{1,1,1,1\}\). Here, \(a_1(b_1)\) and \(a_2(b_2)\) stand for \(A(B)\) atom in upper and lower layer, respectively. \(H_b(\vec{k})\) is the Fourier transform result of effective Hamiltonian \(H_b\) in Eq.1 from real space to momentum space. The driven field will affect the superconductor order parameter. Previously,  it has been proved that superconductivity order parameter $\Delta$ is slightly renormalized in 1D superconductor.\cite{Dong} Similarly, the superconductivity order parameter $\Delta$  in our proposed setup remains unchanged in the limit of \(v_{g}\mathcal{A}/\omega\ll1\) (see App.B for details). For the bilayer graphene system, the parameters \(V\) and \(\Delta\) are in the range of several $meV$. The magnitude of \(m\), determined by the induced light, can be tuned into the same order of \(V\) and \(\Delta\). Without loss of generality, all parameters  including \(m, V\) and \(\Delta\) in Eq.2 are chosen to be positive in the following.

\section{Topological phase diagram}

Firstly, we investigate the topological phase diagram of the proposed system by calculating the topological invariant: Chern number. Since the topological invariant cannot change without closing the bulk energy gap, the gap-closing condition of \(H_{BDG}\) gives out possible topological phase transition points. Based on the diagonalization of \(H_{BDG}\) in Eq.2, we find the bulk energy band closes exactly at the Dirac points. In specific, the energy levels at Dirac points are \(E=\pm\Delta\pm m\pm V\), \(\pm\sqrt{(\Delta\pm m)^2+\gamma^2}\), which means the bulk energy gap closes at two critical conditions: \(m=|\Delta+V|\) and \(m=|\Delta-V|\). Consequently, there may exist three topological phases which can be distinguished by different Chern numbers.

\begin{figure}[h]
\includegraphics [width=\columnwidth, viewport=0 0 944 513, clip]{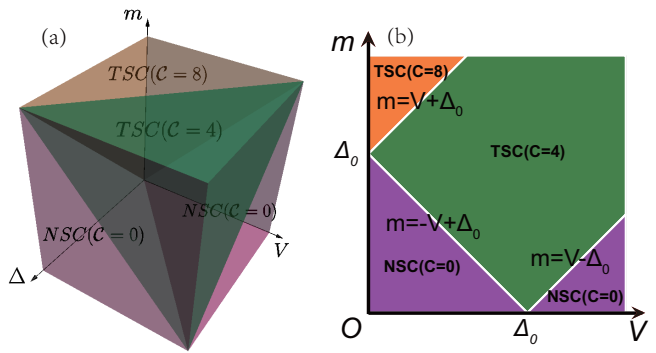}
\caption{(Color online)
Topological phase diagrams of the system.
(a): 3D phase diagram.
There are three regions colored by orange, green and fuchsia.
They have Chern number \(\mathcal{C}=8,4,0\) respectively.
(b): 2D cross section of (a)
in \(V\text{-}m\) plane with fixed \(\Delta_0\).
}
\end{figure}

Based on the calculation result of the Chern number of \(H_{BDG}\)(see App.C), we obtain the 3D topological phase diagram in the first quadrant of \((\Delta,V,m)\) parameter space as shown in Fig.2(a). The orange region is bounded by three planes: \(V=0,\Delta=0\) and \(m=\Delta+V\), which satisfies \(m>\Delta+V\). In App.C, we have calculated the Chern number \(\mathcal{C}=8\) in this region. In other words, the system is a topological superconductor (TSC) with $8$ chiral Floquet Majorana edge states. Similarly, \(m=\Delta+V\), \(m=\Delta-V\) and \(m=V-\Delta\) give out the constraints of the green region. The green region satisfies the condition \(|\Delta-V|<m<|\Delta+V|\) and has a Chern number \(\mathcal{C}=4\), which means this system possesses 4 chiral Floquet Majorana edge states.
Besides, since the Chern number \(\mathcal{C}\) is zero under the condition \(m<|\Delta-V|\),
the two fuchsia regions are both normal superconductor (NSC) with two different sets of boundaries: {\(V=0\), \(m=0\), \(m=\Delta-V\)}
and {\(\Delta=0\), \(m=0\), \(m=V-\Delta\)}.

In experiments, it is common that the superconductor in the proposed system has a fixed superconducting order parameter \(\Delta_0\). Thus, we also draw the cross section of Fig.2(a) in \(V\text{-}m\) plane [see Fig.2(b)]. Mostly, if the positive \(V\) is not equal to \(\Delta_0\), the system can go through three topological phases from $\mathcal{C}=8$ to $\mathcal{C}=4$ and $\mathcal{C}=0$ consecutively by decreasing $m$
as shown in Fig.2(b). That is to say, one can change \(m=v_g^2\mathcal{A}^2/\omega\) by tuning the amplitude \(A\) or frequency \(\omega\) of the induced circularly polarized light, in order to precisely make the system go through three topological phases in experiments with fixed LAF amplitude \(V\) and superconducting order parameter \(\Delta\). Considering the band-width of bilayer graphene, the required frequency of off-resonant light is about $\omega=2500 THz$. For strong light intensity $I=7.5\times10^{11} W/cm^2$, the light induced next nearest-neighbour hopping is around $m=3 meV$. Besides, when \(V\) is equal to \(\Delta_0\) or zero, there exist only two topological phases with the change of \(m\) as shown in Fig.2(b), because one of the two critical conditions is useless now.

\section{Chiral Floquet Majorana edge states}

We further confirm the chiral Floquet Majorana edge states of proposed device with the numerical calculation result obtained from the energy band structure in ribbon geometry. Firstly, \(H_{BDG}\) in Eq.2 can be transformed to real space as:
\begin{eqnarray}
  \tilde{H}_{BDG} & = & H_b+\Delta\sum_{i,\alpha}[a^\dagger_{i,\uparrow,\alpha}a^{\dagger}_{i,\downarrow,\alpha}
                  -a^\dagger_{i,\downarrow,\alpha}a^{\dagger}_{i,\uparrow,\alpha}+H.c.] \nonumber\\
            &   & +\Delta\sum_{i,\alpha}[b^\dagger_{i,\uparrow,\alpha}b^{\dagger}_{i,\downarrow,\alpha}
                  -b^\dagger_{i,\downarrow,\alpha}b^{\dagger}_{i,\uparrow,\alpha}+H.c.]
\end{eqnarray}
where \(H_b\) is given out in Eq.1 and all parameters are same as those in Eq.1 and 2. Then,  considering a bilayer graphene ribbon with periodic boundary in \(x\) direction and open boundary in \(y\) direction, we draw the energy bands by diagonalizing Eq.(3).

\begin{figure}[htbp]
\includegraphics [width=\columnwidth, viewport=0 0 726 366, clip]{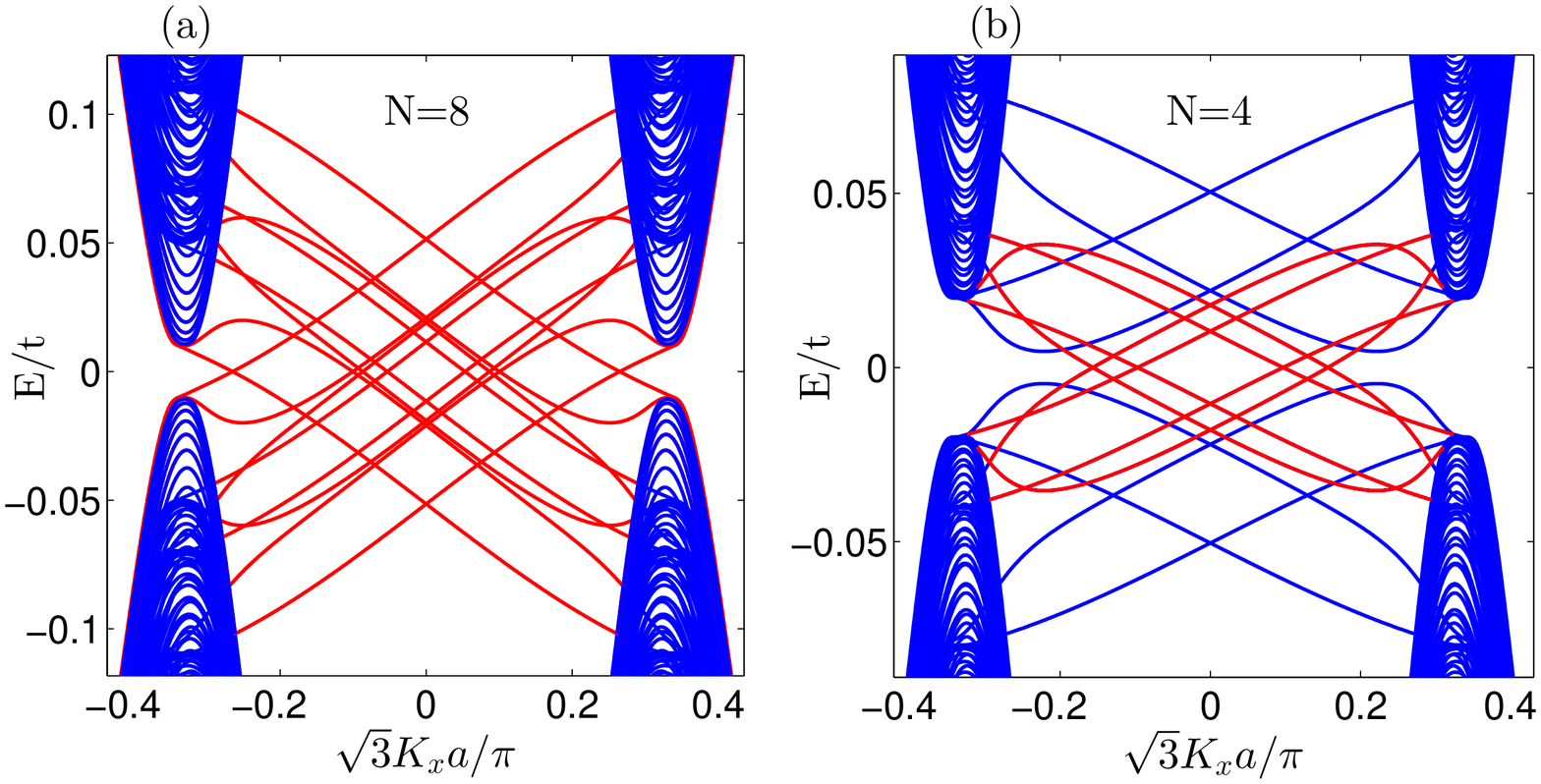}
\includegraphics [width=\columnwidth, viewport=0 0 732 275, clip]{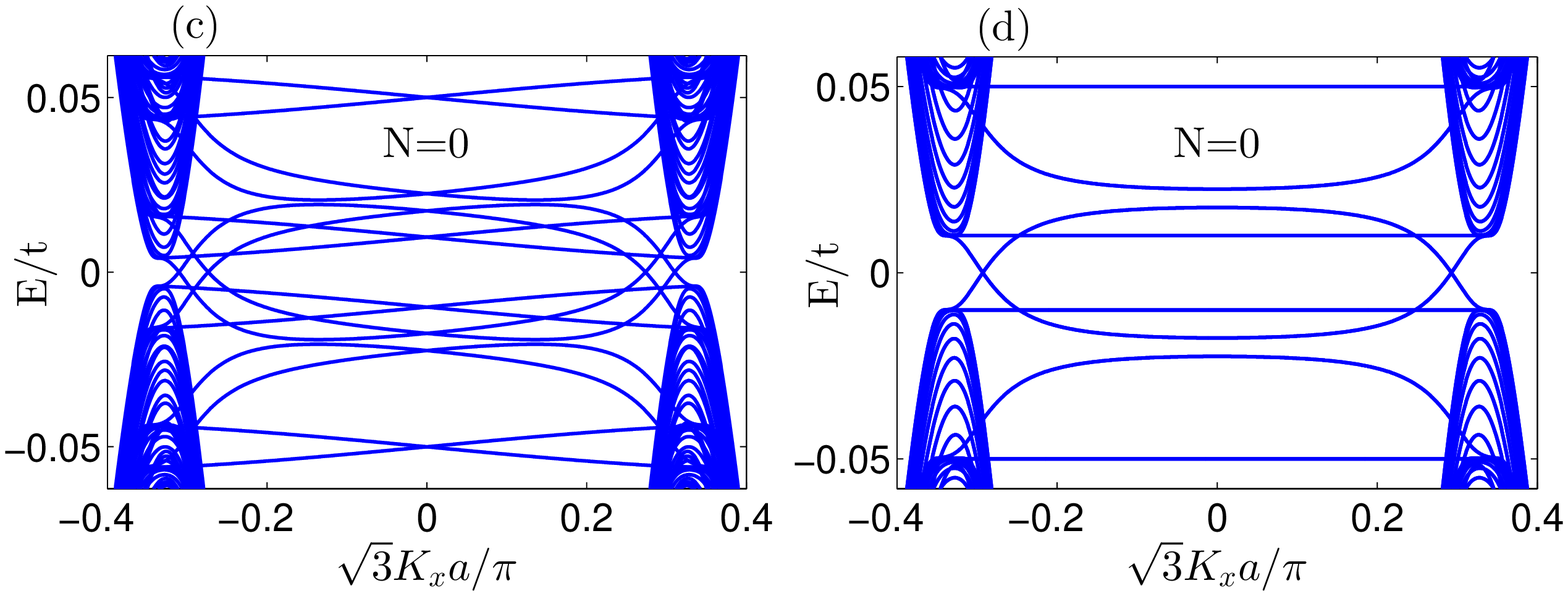}
\caption{(Color online) Energy spectrum of the system with zigzag edges under the condition \(V=0.03,\Delta=0.02\) and \(\gamma=0.3\). The width along \(y\) direction is chosen \(W=240a\) with lattice constant \(a=1\). $N$ is the number of gapless chiral edge states. In (a), \(m=0.06\), the system has eight edge states. In (b), \(m=0.03\), the system has four edge states.
In (c), \(m=0.006\), no edge states exist. In (d), \(m=0\), there is still no edge state. The edge states in (a) and (b) are colored by red.}
\end{figure}

As a heuristic example, in Fig.3, we first study a wide zigzag ribbon with \(W=240a\) and the lattice constant \(a=1\). Similar to the TSC ($\mathcal{C}=8$) region in Sec.III, when parameters satisfy \(m>|V+\Delta|\), there are eight edge states locating inside the bulk band gap [see Fig.3(a)]. Moreover, when \(m\) decreases to the region between \(|V-\Delta|\) and \(|V+\Delta|\), four gapless edge states appear inside the bulk band gap [see Fig.3(b)], corresponding to the aforementioned TSC region with $\mathcal{C}=4$. This topological phase transition indicates that four chiral Floquet Majorana edge states have been annihilated into the bulk states. Finally, if \(m\) is smaller than \(|V-\Delta|\), corresponding to the NSC ($\mathcal{C}=0$) region, there is no chiral edge state inside the bulk band gap [see Fig.3(c)]. This means if the induced circularly polarized light is not strong enough or even absent, the system belongs to the same topologically trivial NSC phase (see Fig.3(c) and 3(d)).

\begin{figure}[htbp]
\includegraphics [width=\columnwidth, viewport=0 0 737 555, clip]{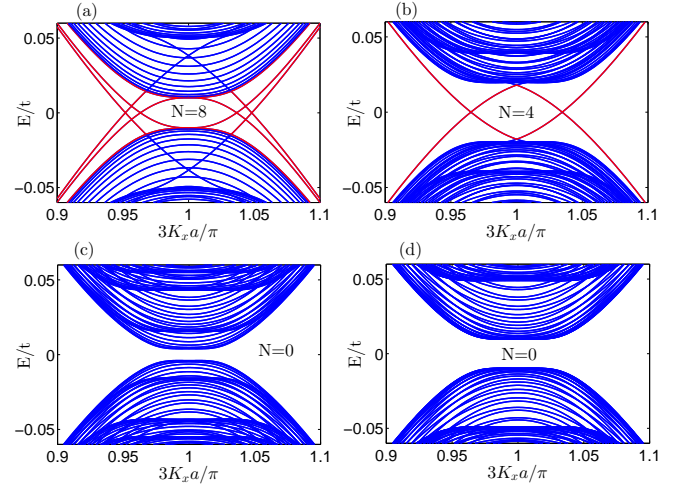}
\caption{(Color online) Energy spectrum of the system with armchair edges under condition \(V=0.03,\Delta=0.02\) and \(\gamma=0.3\). The length along \(y\) direction is chosen \(W=240\sqrt{3}a\) with lattice constant \(a=1\). $N$ is the number of gapless chiral edge states. In (a), \(m=0.06\), the system has four double-degenerate edge states ($N=8$). In (b), \(m=0.03\), the system has two double-degenerate edge states ($N=4$). In (c), \(m=0.006\), no edge states show up. In (d), \(m=0\), there is still no edge state. Similarly, the edge states in (a) and (b) are colored
by red.}
\end{figure}

In order to make the edge states easier to be distinguished,
we also study a wide armchair ribbon with \(W=240\sqrt{3}a\) and the lattice constant \(a=1\) in Fig.4.
The numbers of edge states in different conditions are the same as those in zigzag ribbon. However, due to the undistinguishment of $K$ and $K'$ point in armchair ribbon, the edge states in Fig.4 are double degenerate. Combining Fig.3 and Fig.4, the number of chiral edge states in the wide ribbon is consistent with the Chern number obtained in Sec.III.

\begin{figure}[htbp]
\includegraphics [width=\columnwidth, viewport=0 0 967 334, clip]{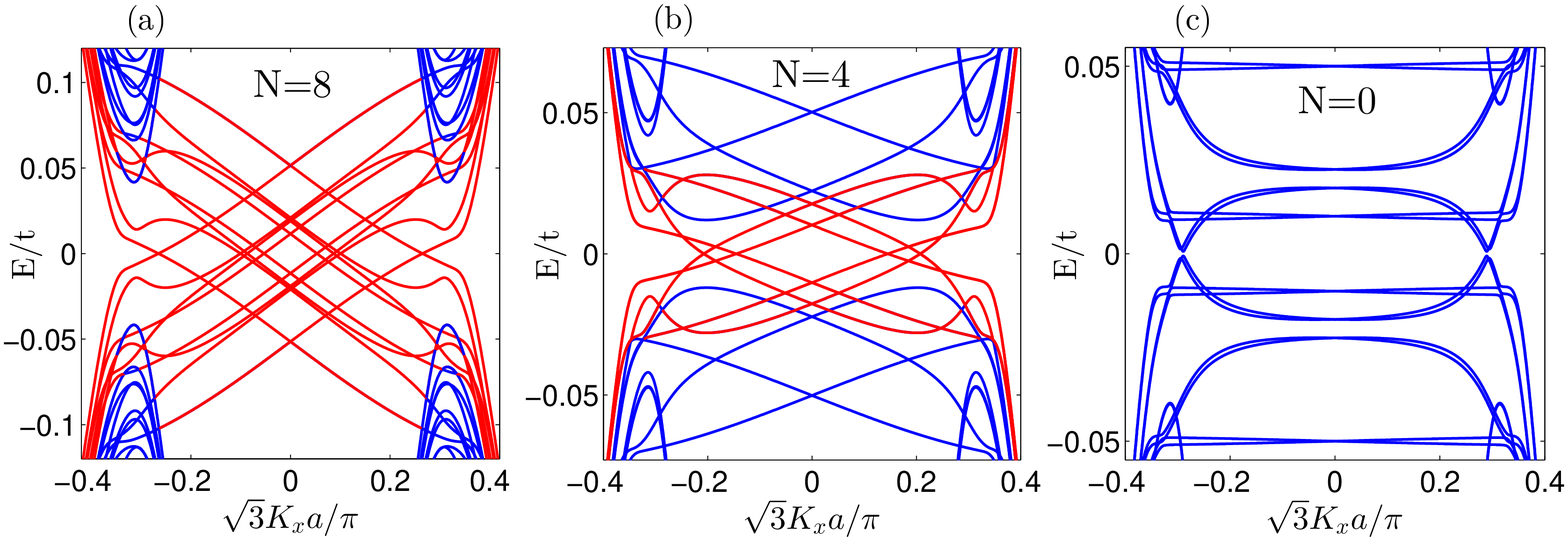}
\includegraphics [width=\columnwidth, viewport=0 0 964 476, clip]{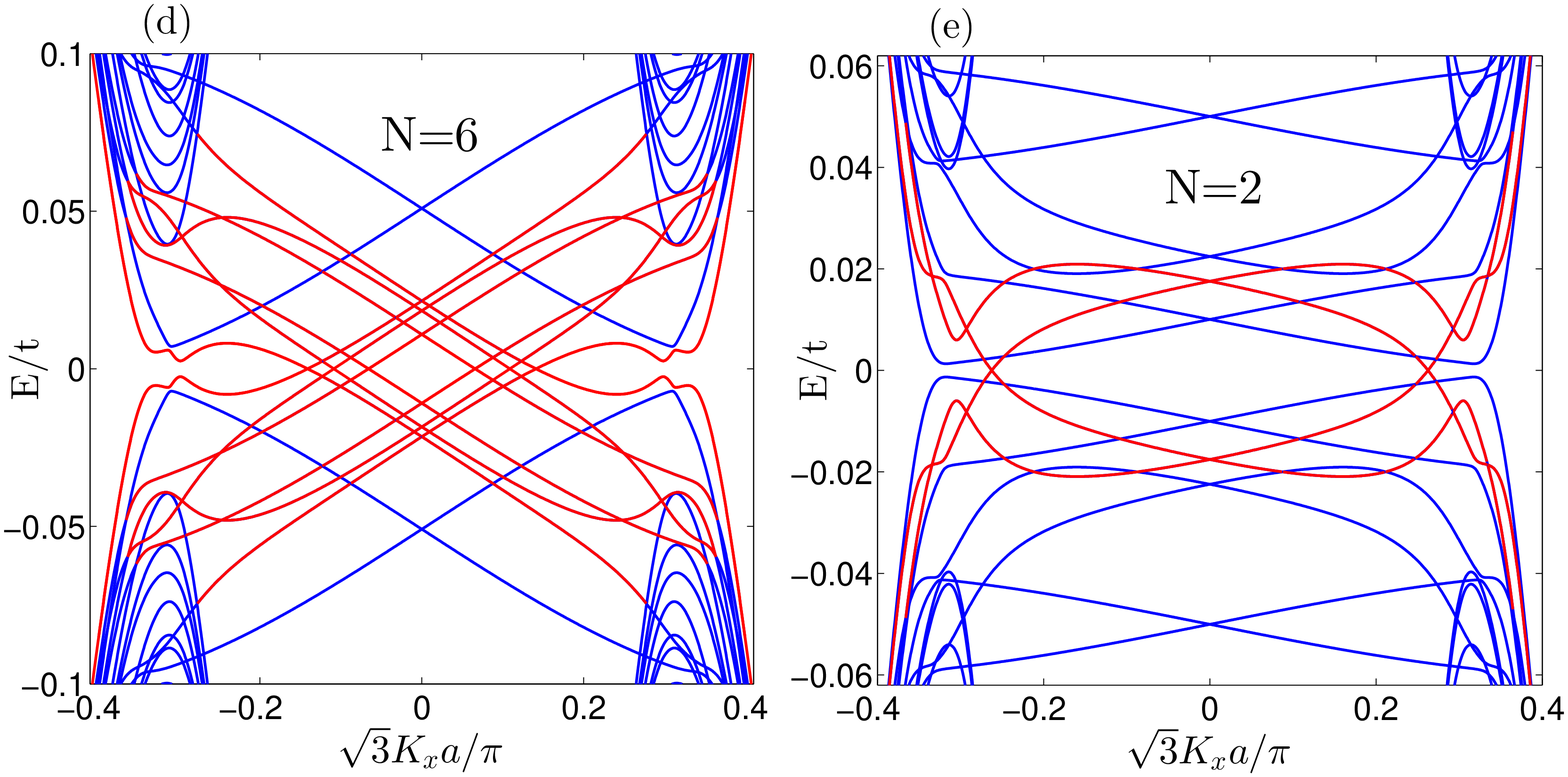}
\caption{(Color online) Energy spectrum of the system with zigzag edges under condition \(V=0.03,\Delta=0.02\) and \(\gamma=0.3\).
The length along \(y\) direction is chosen \(W=60a\) with lattice constant \(a=1\),
which is a quarter of that in Fig.3. $N$ is the number of gapless chiral edge states. In (a), \(m=0.06\), the system has eight edge states.
In (b), \(m=0.02\), the system has four edge states.
In (c), \(m=0.001\), no edge states show up. Those three phases
also exist in Fig.3. However, in (e) and (f), \(m=0.046\) and \(0.009\),
there are six and two edge states, respectively. In order to make the energy spectrum more clearer, the edge states in (a)-(e)
are colored by red.
}
\end{figure}

Surprisingly, in narrow zigzag ribbons, e.g. \(W=60a\), we find two additional parameter regions of $m$ characterized by edge states number six and two, as shown in Fig.5(d) and 5(e), respectively. Besides, the numbers of edges states in Fig.5(a)-5(c) are the same as those in Fig.3(a)-3(c). Similar results can also be found in armchair ribbon, which are not shown here.

The appearance of these two unexpected cases, with six or two edge states in a narrow ribbon, originates from the ``finite size effect''.~\cite{Shen2008,Hua2014} To explain this phenomenon, we firstly utilize Eq.C5-C8 in App.C, which tells us that the characteristic penetration lengths of chiral edge states are: \(\xi_1\sim\frac{2v_f}{|V+m+\Delta|}, \xi_2\sim\frac{2v_f}{|V-m+\Delta|}, \xi_3\sim\frac{2v_f}{|V+m-\Delta|}\)
and \(\xi_4\sim\frac{2v_f}{|V-m-\Delta|}\) in a ribbon. We take parameters in Fig.5 as an example. When \(m\) is large enough compared with \(V+\Delta=0.05\), the ribbon width \(W\) becomes longer than \(\xi_{i=1-4}\), and there exist eight chiral edge states. However, when \(m\) approaches \(V+\Delta=0.05\), the characteristic length \(\xi_2\) becomes very large. Thus, in a narrow ribbon with \(W\)<\(\xi_2\), two chiral edge states with penetration length \(\xi_2\) are coupled with each other, and the number of chiral edge states decreases to six, as shown in Fig.5(d). Then, when \(m\) satisfies \(|V-\Delta|<m<|V+\Delta|\), there exist only four edge states, which are characterized by penetration length \(\xi_3\) and \(\xi_4\). Similarly, if \(m\) approaches to \(|V-\Delta|=0.01\), \(\xi_4\) could become longer than \(W\), and two edge states with penetration length \(\xi_4\) are coupled with each other. Thus, the number of edge states reduces from four to two, as shown in Fig.5(e). At last, when \(m\) becomes small enough compared with \(|V-\Delta|=0.01\), there exists no edge state.


\section{Expanded models and experimental detection}

The aforementioned proposal for generating chiral Floquet Majorana edge states can also be extended to other kinds of hexagonal lattices, such as single-layer graphene and silicene. In particular, considering a single-layer graphene with antiferromagnetic order and shining circularly polarized light, the effective Hamiltonian reads:
\begin{eqnarray}
    H_s & = & -t\sum_{<ij>,s}(a^\dagger_{i,s}b_{j,s}+H.c.) \nonumber\\
      &   & +\frac{im}{3\sqrt{3}}\sum_{\ll ij\gg,s}
       [v_{ij}(a^\dagger_{i,s}a_{j,s}+b^\dagger_{i,s}b_{j,s})] \nonumber\\
      &   & +V\sum_{i,s}[s(a^\dagger_{i,s}a_{i,s}-b^\dagger_{i,s}b_{i,s})]
\end{eqnarray}
where \(a_{i,s}(a^\dagger_{i,s})\) annihilates (creates) an electron with
spin \(s(s=\pm1)\) on site \(\vec{R}_i\) on sublattice
\(A\) (an equivalent definition is used for sublattice \(B\) ). The third term in \(H_s\) denotes the antiferromagnetic coupling, similar to the LAF term in Eq.1. In the presence of strong electron-electron interaction, the neutral graphene manifests an antiferromagnetic phase. On the other hand, when proximating to a superconductor, this system may exhibit topological nontrivial features with chiral Floquet Majorana edge states. Using the similar method as that in bilayer graphene model, we confirm that there still exist two critical conditions corresponding to topological phase transitions: \(m=|V+\Delta|\) and \(m=|V-\Delta|\). Specifically, \(\mathcal{C}=4\) when \(m>|V+\Delta|\), \(\mathcal{C}=2\) when \(|V-\Delta|<m<|V+\Delta|\) and \(\mathcal{C}=0\) when \(m<|V-\Delta|\). Thus, there are three topological phases, in analogy to the case of bilayer graphene.

\begin{figure}[htbp]
\scalebox{0.9}{\includegraphics [width=\columnwidth, viewport=0 0 1365 708, clip]{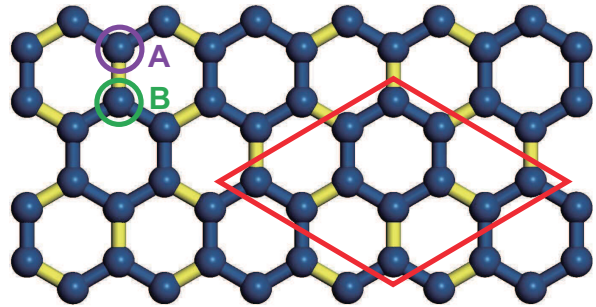}}
\caption{(Color online) The lattice structure of  a \(\sqrt{3}\times\sqrt{3}R\) reconstruction
of monolayer silicene. The nearest-neighbor hopping strength $\tilde{t}$ through the blue bonding is $t-\Delta t$,
while $\tilde{t}$ through the yellow one is $t+\Delta t$. Atoms in the red parallelogram forms
a primitive cell whose side-length is $\sqrt{3}$ times longer than pure silicene. A and B represent
different sublattice.
}
\end{figure}

Moreover, silicene, as another important kind of hexagonal lattice, is the silicon equivalent of graphene,
apart from a relatively large intrinsic Rashba spin orbital term.\cite{Cheng-Cheng Liu1}
In a monolayer silicene, it has been demonstrated both theoretically and experimentally that
there could exist a \(\sqrt{3}\times\sqrt{3}R\) reconstruction,\cite{Lan Chen1,Antoine Fleurence} which is illustrated in Fig.6.
The nearest-neighbor hopping energy through the blue bonding is $t-\Delta t$, while the yellow one
is $t+\Delta t$. Therefore, the primitive cell in monolayer silicene is extended to
$\sqrt{3}\times\sqrt{3}$ times larger than a pure one, which is marked by a red parallelogram
in Fig.6. Considering a \(\sqrt{3}\times\sqrt{3}R\) reconstruction monolayer silicene
system with an antiferromagnetic order and shining circularly polarized light,
the effective Hamiltonian $H_{s}^{Si}$ can be written as:
\begin{eqnarray}
    H_s^{Si} & = & -\sum_{<ij>,s}(\tilde{t}_{i,j}a^\dagger_{i,s}b_{j,s}+H.c.) \nonumber\\
      &   & -it_{SO}\sum_{\ll ij\gg ss'}\mu_{ij}[a_{i,s}^{\dagger}(\vec{\sigma}\times\hat{\vec{d}}_{ij})^z_{ss'}a_{j,s'}\nonumber \\
      &   &+b_{i,s}^{\dagger}(\vec{\sigma}\times\hat{\vec{d}}_{ij})^z_{ss'}b_{j,s'}] \nonumber\\
      &   & +\frac{im}{3\sqrt{3}}\sum_{\ll ij\gg,s}
       [v_{ij}(a^\dagger_{i,s}a_{j,s}+b^\dagger_{i,s}b_{j,s})] \nonumber\\
      &   & +V\sum_{i,s}[s(a^\dagger_{i,s}a_{i,s}-b^\dagger_{i,s}b_{i,s})]
\end{eqnarray}
where \(a_{i,s}(a^\dagger_{i,s})\), \(b_{i,s}(b^\dagger_{i,s})\) and \(v_{ij}\)
have the same meanings as those in Eq.4.
The first term stands for the nearest-neighbor hopping, and the definition of $\tilde{t}_{ij}$
can be found in Fig.6. Specifically, in blue bonding, $\tilde{t}_{ij}=t-\Delta t$,
while $\tilde{t}_{ij}=t+\Delta t$ in yellow. $\tilde{t}_{ij}$ is the source
that folds valleys $K$ and $K'$ into the $\Gamma$ point, and results in intervalley scattering.
The second term is the intrinsic Rashba spin orbital term
with $t_{SO}$ the corresponding strengths. $\hat{\vec{d}}_{ij}=\vec{d}_{ij}/|\vec{d}_{ij}|$,
where $\vec{d_{ij}}$ represents a vector from site $j$ to $i$, and $\mu_{ij}=\pm1$
for an A or B site. The third and the last term denote respectively the effect of circularly polarized light
and antiferromagnetism.

\begin{figure}[htbp]
\includegraphics [width=\columnwidth, viewport=0 0 730 375, clip]{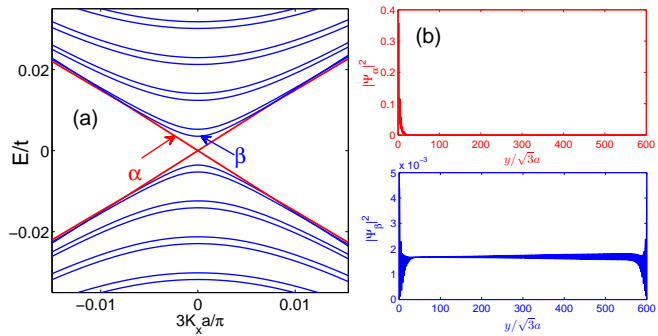}
\caption{(Color online) (a): Energy spectrum of \(\sqrt{3}\times\sqrt{3}R\) reconstruction of
monolayer silicene with armchair edges under the condition $\Delta t=0.098, t_{SO}=0.5, m=0.2, V=0.1, \Delta=0.14$.
The width along $y$ direction is chosen as $W = 600\sqrt{3}a$ with
lattice constant $a = 1$. There is only one edge state marked by $\alpha$, and the second lowest state
is marked by $\beta$.
(b): The distribution of wave functions $|\Psi_{\alpha}|^2$ and $|\Psi_{\beta}|^2$ colored by red and blue,
respectively.
}
\end{figure}

Considering the similarity of silicene and graphene, our FMFs proposal is also applicable to the silicene system.
But unlike the graphene system, owing to the \(\sqrt{3}\times\sqrt{3}R\) reconstruction and the intrinsic Rashba spin orbital term in Eq.5,
there may exist other topological phases with odd chiral edge states.
Notably, the realization of odd numbers of FMFs is important in the realization of nontrivial braiding statistics, which is
important for the quantum topological computation.
In Fig.7, we study an armchair ribbon of the \(\sqrt{3}\times\sqrt{3}R\) reconstruction monolayer silicene.
The width of the ribbon is chosen as $W=600\sqrt{3}a$ with lattice constant $a=1$.
In Fig.7(a), it is clear that under the condition $\Delta t=0.098, t_{SO}=0.5, m=0.2, V=0.1, \Delta=0.14$,
there exists only one edge state, marked with $\alpha$. In order to confirm that the second lowest
state $\beta$ belongs to the bulk bands, the wave function distributions $|\Psi_{\alpha(\beta)}|^2$
of states $\alpha$ and $\beta$ are also drawn in Fig.7(b). We find that
$|\Psi_{\alpha}|^2$ distributes mostly near the boundary, while $|\Psi_{\beta}|^2$
distributes inside the ribbon, which means state $\beta$ is a bulk state. Thus, the
\(\sqrt{3}\times\sqrt{3}R\) reconstruction monolayer silicene system possesses only one
chiral edge state hosting odd number of FMFs.

\begin{figure}[h]
\includegraphics [width=\columnwidth, viewport=0 0 1314 478, clip]{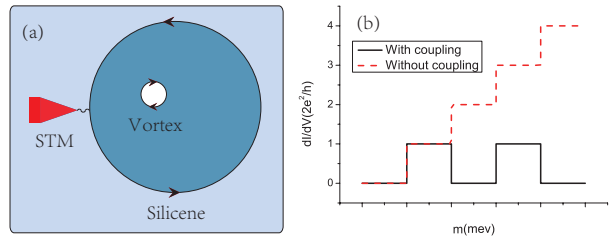}
\caption{(Color online)
Experimental expectation.
(a): Schematic of the apparatus. A round piece of \(\sqrt{3}\times\sqrt{3}R\) reconstructed silicene is put in proximity to a superconductor with induced light. Considering a $\frac{hc}{2e}$ flux is induced in the system, a STM tip is used to probe the \(dI/dV\) characteristics of Majorana zero modes.
(b): Zero energy \(dI/dV\) peak value. When the Majorana zero modes in vortices decouple, the peak value increases to be integer times of $\frac{2e^{2}}{h}$. When the Majorana zero modes in vortices couple together, the peak value shows oscillation behavior.
}
\end{figure}

The experimental detection of Majorana Fermions, which do not transport charge, is impeded by the difficulties to measure spin and heat transport. Thus, precise experimental setups are needed for detecting those states. The topological phases and topological critical points in our proposed systems can be detected by the zero bias \(dI/dV\) characteristic curves of scanning tunneling spectroscopy.~\cite{Law2009}

Firstly, taking silicence system as an example, when a flux $\frac{hc}{2e}$ is induced in the system, the Majorana zero modes emerge at the inner and outer edges.~\cite{Green2000} When the STM tip couples to the Majorana zero modes through electron tunneling, the resonant Andreev reflections can induce zero bias peak in the \(dI/dV\) characteristic curves.~\cite{Law2009} Specifically, as shown in Fig.8, when the Majorana zero modes in vortices do not couple to each other, the peak value increases to be integer times of $\frac{2e^{2}}{h}$ when the topological phase transition happens. In a sharp contrast, when the Majorana zero modes in vortices couple together, the peak value shows oscillation behavior at the topological phase transition points.

Moreover, we consider the bilayer graphene system with three topological phases characterized by Chern numbers 8, 4 and 0. There are two cases for the bilayer graphene system. (i) If the coupling between chiral Majorana edge states can be neglected, the experimental detection setup is similar to the silicence system.  Considering a flux $\frac{hc}{2e}$ induced in the system, the Majorana zero modes emerge at the vortex boundary. The zero bias \(dI/dV\) peak value is related to bulk Chern number by the relation of $G=\mathcal{C}\times\ensuremath{\frac{2e^{2}}{h}}$. (ii) If the coupling between edge states cannot be neglected, the degeneracy of edge states is lifted, thus inducing flux $\frac{hc}{2e}$ cannot guarantee the existence of zero modes. However, considering the induced flux is tuned in the range of $\left[0\text{,\ensuremath{\frac{hc}{e}}}\right)$, the zero modes still turn up at certain flux value. For this case, during the induced flux is tuned from $0$ to $\frac{hc}{e}$, the zero bias \(dI/dV\) characteristic curve shows oscillation behavior as function of the flux, with $\mathcal{C}$ times peaks at value $\frac{2e^{2}}{h}$, where $\mathcal{C}$ is the bulk Chern number. One the other hand, the local density of edge states changes when the Chern number is changed, and leads to observable effects in the tunneling conductance in normal-Floquet topological superconductor junction.\cite{Tanaka} The aforementioned experimental detection methods can also be applied to the monolayer graphene system.

\section{Conclusion}

In conclusion, when hexagonal lattice layer with circularly polarized light is put in proximity to an s-wave superconductor, there exist topological phases with chiral Floquet Majorana edge states. Specifically, in the proposed bilayer graphene system with layer-antiferromagnetic term, there are three topological phases with the Chern number 8, 4, 0 by tuning the amplitude \(A\) or the frequency \(\omega\) of the induced light. The same proposal can also be extended to other kinds of hexagonal lattices, i.e. monolayer graphene and silicene.
Notably, the \(\sqrt{3}\times\sqrt{3}R\) reconstruction monolayer silicene system can host odd numbers of FMFs, which is important
for quantum topological computation. To sum up, by accurately adjusting the frequency and the amplitude of induced circularly polarized light, one can precisely tune the number of Majorana edge states, achieve Floquet Majorana fermions, and realize topological phase transitions in hexagonal lattices.
In experiments, those Floquet Majorana fermions can be characterized by zero bias \(dI/dV\) peak using scanning tunneling spectroscopy.

\section*{ACKNOWLEDGMENTS}
We thank the insightful discussions with Qing-feng Sun, Fa Wang, Jie Liu and Pei Wang. This work is financially supported
by MOST of China (2012CB821402), NSF-China under Grants No. 91221302 and No. 1134219, China Post-doctoral Science Foundation under Grant No. 2012M520099,
and NSFC of Jiangsu province SBK201340278.

\begin{appendix}
\section{BILAYER GRAPHENE EFFECTIVE HAMILTONIAN}

When a beam of circularly polarized light with vector potential
\(\vec{A}(t)=A(\pm sin(\omega t), cos(\omega t))\), where \(+1\) is for right circulation
of light and \(-1\) for left circulation, is applied perpendicularly to a
bilayer graphene without other effects, the Hamiltonian \(H(t)\) can be written as:
\begin{eqnarray}
   H(t)&=&-t\sum_{<ij>,s,\alpha}(e^{iA_{ij}(t)}a^\dagger_{i,s,\alpha}b_{j,s,\alpha}+H.c.) \nonumber\\
       & &-\gamma\sum_{i,s}(a^\dagger_{i,s,1}a_{i,s,2}+H.c.)
\end{eqnarray}
where \(A_{ij}(t)=e/\hbar(\vec{r}_j-\vec{r}_i)\cdot\vec{A}(t)\), with \(\vec{r}_i\)
being the coordinates of the lattice site \(i\), \(\alpha=1,2\) the layer index,
\(t\) the hopping amplitude of the nearest electrons
in the same layer, \(\gamma\) the hopping amplitude between electron $A_1$ and $A_2$ in different layers,
and \(s\) the spins of electrons, as shown in Fig.1(b).

In this way, the Hamiltonian near the Dirac point \(K=-\frac{4\pi}{3\sqrt{3}a}\vec{e}_x\) is
\begin{eqnarray}
     H_K(t)&=&v_ga I_2\otimes(\sigma_x\mathcal{K}_x+\sigma_y\mathcal{K}_y)-\frac{\gamma}{2}(\sigma_x\otimes\sigma_x-\sigma_y\otimes\sigma_y) \nonumber\\
\end{eqnarray}
where \(v_g=\frac{3}{2}t\), \(\mathcal{K}_{x(y)}=k_{x(y)}+\frac{eA_{x(y)}}{\hbar}\),
\(k_x\) and \(k_y\) are momenta measured from the Dirac point \(K=-\frac{4\pi}{3\sqrt{3}a}\vec{e}_x\),
\(I_2=diag\{1,1\}\), and \(\sigma_i\) is Pauli matrix.

In Ref.[\onlinecite{Takuya Kitagawa2, Mark S. Rudner}], we know that the effective hamiltonian of the system
in off-resonant condition is defined as:
\begin{eqnarray}
   H_{eff}=\frac{i}{T}log(U)
\end{eqnarray}
where \(U=\mathcal{T}exp(-i\int_0^T H(t)\dif t)\) and \(\mathcal{T}\) is
the time-ordering operator.

In the limit of \(\mathcal{A}^2\ll1\), with \(\mathcal{A}=eAa/\hbar\),
\begin{eqnarray}
   U&=&\mathcal{T}exp(-i\int_0^T H(t)\dif t)  \nonumber\\
    &=&\mathcal{T}\sum_{n=0}^\infty\frac{[-i\int_0^T H(t)\dif t]^n}{n!}  \nonumber\\
    &\approx&\mathcal{T}[1-i\int_0^T H(t)\dif t-\frac{\int_0^T\dif t_1\int_0^T\dif t_2 H(t_1)H(t_2)}{2}]  \nonumber\\
    &=&1-iTH_0  \nonumber\\
    & &-\frac{H_0^2T^2}{2}-[H_1,H_{-1}]\frac{T}{i\omega}-[H_0,H_1-H_{-1}]\frac{T}{i\omega}
\end{eqnarray}
where \(H_n=\frac{1}{T}\int_0^T H(t)e^{in\omega t}\dif t\).
And the effective Hamiltonian can be deduced as:
\begin{eqnarray}
   H_{eff}&=&\frac{i}{T}log(U)  \nonumber\\
          &\approx&H_0-\frac{[H_1,H_{-1}]}{\omega}-\frac{[H_0,H_1-H_{-1}]}{\omega}
\end{eqnarray}

After applying \(H_K(t)\) into the definition of \(H_n\), it is straightforward to obtain:
\begin{eqnarray}
   [H^K_{1},H^K_{-1}]&=&\mp v_g^2\mathcal{A}^2I_2\otimes\sigma_z \\
   {[H^K_{0},H^K_{1}-H^K_{-1}]}&=&\pm v_g\mathcal{A}(2v_gaK_yI_2+\gamma\sigma_y)\otimes\sigma_z \nonumber\\
\end{eqnarray}

Since \(v_g\mathcal{A}/\omega\ll1\) and the system is considered near
Dirac point \(K\), it is obvious that
\begin{eqnarray}
   \frac{[H^K_{0},H^K_{1}-H^K_{-1}]}{\omega}\ll H^K_{0}-\frac{[H^K_{1},H^K_{-1}]}{\omega}
\end{eqnarray}
Consequently, \(H_{Keff}\) is quite simple now:
\begin{eqnarray}
   H_{Keff}&\approx&H^K_{0}-\frac{[H^K_{1},H^K_{-1}]}{\omega}  \nonumber\\
          &=&v_ga I_2\otimes(\sigma_xk_x+\sigma_yk_y)-\frac{\gamma}{2}(\sigma_x\otimes\sigma_x-\sigma_y\otimes\sigma_y) \nonumber\\
          &&\pm \frac{v_g^2\mathcal{A}^2}{\omega}I_2\otimes\sigma_z
\end{eqnarray}
Just as the same process above, the Hamiltonian \(H_{eff}\) near the other
Dirac point \(K'=\frac{4\pi}{3\sqrt{3}a}\vec{e}_x\) can be written as:
\begin{eqnarray}
   H_{K'eff}&\approx&H^{K'}_{0}-\frac{[H^{K'}_{1},H^{K'}_{-1}]}{\omega}  \nonumber\\
            &=&v_ga I_2\otimes(-\sigma_xk'_x+\sigma_yk'_y)-\frac{\gamma}{2}(\sigma_x\otimes\sigma_x-\sigma_y\otimes\sigma_y) \nonumber\\
            &&\mp \frac{v_g^2\mathcal{A}^2}{\omega}I_2\otimes\sigma_z
\end{eqnarray}
where \(k'_x\) and \(k'_y\) are momenta measured from the Dirac point \(K'=\frac{4\pi}{3\sqrt{3}a}\vec{e}_x\),

Therefore, in real space, the second term \(-\frac{[H_{1},H_{-1}]}{\omega}\) in \(H_{eff}\) can be illustrated
as the second-neighbor hopping \(\sum_{\ll ij\gg} v_{ij}c^{\dagger}_ic_j\)
in each layer, with \(v_{ij}=1\) for hopping clockwise in one hexagonal lattice and \(-1\)
for hopping anti-clockwise. As a result, the effective Hamiltonian \(H_{eff}\) for
bilayer graphene with circularly polarized light perpendicular to its plane is:
\begin{eqnarray}
   H_{eff}&=&-t\sum_{<ij>,s,\alpha}(a^\dagger_{i,s,\alpha}b_{j,s,\alpha}+H.c.)  \nonumber\\
          & &-\gamma\sum_{i,s}(a^\dagger_{i,s,1}a_{i,s,2}+H.c.)  \nonumber\\
          & &\pm\frac{im}{3\sqrt{3}}\sum_{\ll ij\gg,s,\alpha}
             (v_{ij}a^\dagger_{i,s,\alpha}a_{j,s,\alpha}+v_{ij}b^\dagger_{i,s,\alpha}b_{j,s,\alpha}) \nonumber\\
\end{eqnarray}
where \(m=v_g^2\mathcal{A}^2/\omega\), and other parameters mean the same as what we have
mentioned before.

At last, we must emphasize that \(H_{eff}\) in Eq.A11 in this appendix is valid near
Dirac points, thus all other equations deduced from it in this paper must satisfy the same condition.
However, since the topological properties remain unchanged whether
the Hamiltonian we use is valid in whole Brillouin zone or just around Dirac points,
it is fine to apply the effective Hamiltonian \(H_{eff}\) and its deductions to the calculation of energy bands and
Chern number. Therefore, we can continue
discussing the topological phase transition of this system
in this paper.

\section{THE EFFECT OF LIGHT TO SUPERCONDUCTIVITY}

The total time-dependent Hamiltonian of bilayer graphene with circularly polarized light reads:
\begin{eqnarray}
    H_{total}&=&-t\sum_{<ij>,s,\alpha}(e^{iA_{ij}(t)}a^\dagger_{i,s,\alpha}b_{j,s,\alpha}+H.c.) \nonumber\\
             & &-\gamma\sum_{i,s}(a^\dagger_{i,s,1}a_{i,s,2}+H.c.) \nonumber\\
       & & +V\sum_{i,s}\ [s(b^\dagger_{i,s,1}b_{i,s,1}-b^\dagger_{i,s,2}b_{i,s,2})]\nonumber\\
       & & +\Delta\sum_{i,\alpha}[a^\dagger_{i,\uparrow,\alpha}a^{\dagger}_{i,\downarrow,\alpha}
                  -a^\dagger_{i,\downarrow,\alpha}a^{\dagger}_{i,\uparrow,\alpha} \nonumber\\
       & &    +b^\dagger_{i,\uparrow,\alpha}b^{\dagger}_{i,\downarrow,\alpha}
                  -b^\dagger_{i,\downarrow,\alpha}b^{\dagger}_{i,\uparrow,\alpha}+H.c.]
\end{eqnarray}
Utilizing the similar derivation in App.A, Eq.B1 can also be further deduced to:
\begin{eqnarray}
   H_{eff}&\approx&H_0-\frac{[H_1,H_{-1}]}{\omega}-\frac{[H_0,H_1-H_{-1}]}{\omega}
\end{eqnarray}
where \(H_n=\frac{1}{T}\int_0^T H(t)e^{in\omega t}\dif t\), and Eq.B2 is exactly the same as Eq.A5 in App.A. Since only the first term of Eq.B1 contains time variable $t$, the superconducting term makes no contribution to the $H_1$ and $H_{-1}$.
Thus, the commutation $[H_1,H_{-1}]$ only gives out the block-diagonal terms in the effective Hamiltonian.
Moreover, due to the condition \(v_g\mathcal{A}/\omega\ll1\), Eq.A8 is still valid, which means
the commutation $[H_0,H_1-H_{-1}]$ can be ignored.
Therefore, the superconducting term in effective Hamiltonian $H_{eff}$ just appears in $H_0$, and it should remain unchanged
under the condition \(v_g\mathcal{A}/\omega\ll1\).

\section{CALCULATION OF THE CHERN NUMBER}

In Ref.[\onlinecite{A. H. Castro Neto}], the effective low energy Hamiltonian for
bilayer graphene without LAF and induced light can be written as:
\begin{eqnarray}
          \mathcal{H}(k)=\phi_{\vec{k}}^\dagger
   \left(
    \begin{array}{cc}
     0 & \frac{v_F^2}{\gamma}k_-^2\\
     \frac{v_F^2}{\gamma}k_+^2 & 0\\
    \end{array}
   \right)\phi_{\vec{k}}
\end{eqnarray}
where \(\phi_{\vec{k}}=(b_{1k}, b_{2k})^T\), \(k_\pm=k_x\pm ik_y\)
and \(v_F=3at/2\). The low energy physics is mainly determined by wavefunctions located
at \(B_1\) and \(B_2\) atoms, while \(A_1\) and \(A_2\) denote the high energy physics
for \(E>\gamma\).

In the following part, we give out the low energy Hamiltonian considering the LAF term, the circularly polarized light term and superconducting pairing. The high energy atoms \(A_1\) and \(A_2\) are omitted. And the basis can be given as:
\((b_{1\vec{k}\uparrow}, b_{2\vec{k}\uparrow}, -b_{2-\vec{k}\downarrow}^\dagger, b_{1-\vec{k}\downarrow}^\dagger,
b_{1-\vec{k}\uparrow}, b_{2-\vec{k}\uparrow}, -b_{2\vec{k}\downarrow}^\dagger, b_{1\vec{k}\downarrow}^\dagger)^T\)
Moreover, there also exits another half of the basis:
\((b_{1\vec{k}\downarrow}, b_{2\vec{k}\downarrow}, -b_{2-\vec{k}\uparrow}^\dagger, b_{1-\vec{k}\uparrow}^\dagger,
b_{1-\vec{k}\downarrow}, b_{2-\vec{k}\downarrow}, -b_{2\vec{k}\uparrow}^\dagger, b_{1\vec{k}\uparrow}^\dagger)^T\).
The whole basis consisting of sixteen elements is also called Beenakker's Notation.

Considering the first half basis, the first building block can be written as:
\begin{eqnarray}
   \mathcal{H}_1=
   \left(
    \begin{array}{cccc}
     V+m & \lambda k_-^2 & 0 & \Delta\\
     \lambda k_+^2 & -V-m & -\Delta & 0\\
     0 & -\Delta & -V-m & \lambda k_-^2\\
     \Delta & 0 & \lambda k_+^2 & V+m\\
    \end{array}
   \right)
\end{eqnarray}
with basis \((b_{1\vec{k}\uparrow}, b_{2\vec{k}\uparrow}, -b_{2-\vec{k}\downarrow}^\dagger, b_{1-\vec{k}\downarrow}^\dagger)\),
where \(\lambda=v_F^2/\gamma=(\frac{3}{2}at)^2/\gamma\).
And the second block can be written as:
\begin{eqnarray}
   \mathcal{H}_2=
   \left(
    \begin{array}{cccc}
     V-m & \lambda k_+^2 & 0 & \Delta\\
     \lambda k_-^2 & -V+m & -\Delta & 0\\
     0 & -\Delta & -V+m & \lambda k_+^2\\
     \Delta & 0 & \lambda k_-^2 & V-m\\
    \end{array}
   \right)
\end{eqnarray}
with basis \((b_{1-\vec{k}\uparrow}, b_{2-\vec{k}\uparrow}, -b_{2\vec{k}\downarrow}^\dagger, b_{1\vec{k}\downarrow}^\dagger)\).

Then, an unitary transform \(\tilde{\mathcal{H}}=T^\dagger\mathcal{H}T\) can be done
by:
\begin{eqnarray}
    T=\frac{1}{\sqrt{2}}
   \left(
    \begin{array}{cccc}
     1 & 0 & 0 & 1\\
     0 & 1 & 1 & 0\\
     0 & -1 & 1 & 0\\
     -1 & 0 & 0 & 1\\
    \end{array}
   \right)
\end{eqnarray}
Therefore, \(\tilde{\mathcal{H}}_1\) can be diagonalized as \(diag\{\tilde{\mathcal{H}}_{1A}, \tilde{\mathcal{H}}_{1B}\}\),
where
\begin{eqnarray}
   \tilde{\mathcal{H}}_{1A}&=&
   \left(
    \begin{array}{cc}
     V+m-\Delta & \lambda k_-^2\\
     \lambda k_+^2 & -(V+m-\Delta)
    \end{array}
   \right)\\
   \tilde{\mathcal{H}}_{1B}&=&
   \left(
    \begin{array}{cc}
     -(V+m+\Delta) & \lambda k_-^2\\
     \lambda k_+^2 & V+m+\Delta\\
    \end{array}
   \right)
\end{eqnarray}
and \(\tilde{\mathcal{H}}_2\) can be diagonalized as \(diag\{\tilde{\mathcal{H}}_{2A}, \tilde{\mathcal{H}}_{2B}\}\),
where
\begin{eqnarray}
   \tilde{\mathcal{H}}_{2A}&=&
   \left(
    \begin{array}{cc}
     V-m-\Delta & \lambda k_+^2\\
     \lambda k_-^2 & -(V-m-\Delta)
    \end{array}
   \right)\\
   \tilde{\mathcal{H}}_{2B}&=&
   \left(
    \begin{array}{cc}
     -(V-m+\Delta) & \lambda k_+^2\\
     \lambda k_-^2 & V-m+\Delta\\
    \end{array}
   \right)
\end{eqnarray}

According to Berry phase curvature, the Chern number for block 1 is:
\begin{eqnarray}
   \mathcal{C}_{1A}&=&sgn[m+(V-\Delta)]\\
   \mathcal{C}_{1B}&=&sgn[m-(V-\Delta)]
\end{eqnarray}
Combining \(\mathcal{C}_{1A}\) and \(\mathcal{C}_{1B}\), we get:
\begin{eqnarray}
   \mathcal{C}_{1}&=&sgn[m+(V-\Delta)]+sgn[m-(V-\Delta)]  \nonumber\\
                  &=&
                  \left\{
                  \begin{array}{cc}
                  +2 & m>|V-\Delta|\\
                  0  & -|V-\Delta|<m<|V-\Delta|\\
                  -2 & m<-|V-\Delta|
\end{array}
\right.
\end{eqnarray}
In the same way, the Chern number for block 2 can be deduced as:
\begin{eqnarray}
   \mathcal{C}_{2}&=&sgn[m+(V+\Delta)]+sgn[m-(V+\Delta)]  \nonumber\\
                  &=&
                  \left\{
                  \begin{array}{cc}
                  +2 & m>|V+\Delta|\\
                  0  & -|V+\Delta|<m<|V+\Delta|\\
                  -2 & m<-|V+\Delta|
\end{array}
\right.
\end{eqnarray}
Consequently, the whole Chern number can be obtained:
\begin{eqnarray}
       \mathcal{C}_I&=&
                  \left\{
                  \begin{array}{cc}
                  +4 & m>|V+\Delta|\\
                  +2  & -|V-\Delta|<m<|V+\Delta|\\
                  0  & m<|V-\Delta|
\end{array}
\right.
\end{eqnarray}

Moreover, the Chern number of the second half of the basis can also be calculated in the similar method,
and we obtain the same result as \(\mathcal{C}\). Therefore, the total Chern number of the proposed
system is:
\begin{eqnarray}
       \mathcal{C}&=&
                  \left\{
                  \begin{array}{cc}
                  +8 & m>|V+\Delta|\\
                  +4  & -|V-\Delta|<m<|V+\Delta|\\
                  0  & m<|V-\Delta|
\end{array}
\right.
\end{eqnarray}

From the above criterion, we can immediately obtain the phase diagram in Sec.III and uncover the physical picture of
topological phase transitions.
\end{appendix}

\end{document}